\documentstyle[12pt]{article}

\begin{document}

\title{How We Got Into The Dark Matter Fix And How We 
Can Get Out\footnote{astro-ph/9807122, July 13, 1998. 
To appear in the Proceedings of 
"PASCOS 98", the 6th Int. Symp. on Particles, Strings 
and Cosmology, Northeastern Univ., March 1998.}}


\author{\normalsize{Philip D. Mannheim} \\
\normalsize{Department of Physics,
University of Connecticut, Storrs, CT 06269} \\ 
\normalsize{mannheim@uconnvm.uconn.edu} \\}

\date{}

\maketitle

\begin{abstract}
The recent discovery of cosmic repulsion represents a major
challenge to standard gravity, adding an apparent missing energy 
problem to its still not yet adequately resolved missing mass one, 
while simultaneously requiring the current universe to be a very 
special, fine-tuned one. We suggest that the resolution of these 
difficulties lies not in the existence of more and more exotic 
gravitational sources, but rather in the lack of reliability of the 
extrapolation of standard gravity from its solar system origins to 
the altogether larger distance scales associated with cosmology. We 
suggest a very different such extrapolation, namely that associated 
with conformal gravity, to find that all of these issues are then 
readily and naturally resolved.
\end{abstract}

At the present time the status of the standard Newton-Einstein 
gravitational theory is somewhat paradoxical. For a theory whose
correctness is so overwhelmingly agreed upon that the entire issue 
is even considered long since settled, it is extremely disquieting
that the standard theory seems to so frequently need fixing up 
as new data come on line. Indeed, not only has normal, visible matter
long since been demoted by non-baryonic dark matter into being only a 
minor contributor to the energy density of the universe, but, in light of 
the recent detection \cite{Garnavich1998,Perlmutter1998} of an apparent 
cosmic repulsion, this still highly elusive dark 
matter itself (be it hot or cold) has, in its stead, now also been demoted 
by yet more exotic gravitational sources such as a cosmological constant or 
a possible quintessence (non-luminous matter with negative pressure). Thus 
given its presumed correctness, it is perplexing that the standard theory so 
often finds itself in need of remodeling; with this repeated 
remodeling (which is always done only after the fact) entailing, after 
generations of work, that the cosmology community apparently has only a 
rather vague notion as to what the universe is predominantly made of.

Even more disturbing than this is the fact that once the familiar current 
era cosmological matter density $\Omega_M$ is in fact (as now 
definitively shown by the cosmic repulsion data) found to be less than one 
(but apparently still found large enough to still require overwhelming 
amounts of non-baryonic dark matter), the standard macroscopic 
Einstein-Friedmann cosmological evolution equation then requires some new 
fine tuning above and beyond that already provided by the popular flat 
inflationary universe paradigm. Specifically, either the spatial curvature 
$k$ of the universe is negative, so that we are then once again back to the 
original fine tuning flatness problem, or some exotic gravitational source 
of energy density $\Omega_{EX}$ then exists which precisely brings 
$\Omega_{M}+\Omega_{EX}$ back to the flat universe value of one, 
with $\Omega_{EX}$ then needing to have been fine-tuned in the early 
universe to incredible accuracy (possibly by as much as 60 orders of 
magnitude) in order for it to actually be of order one today. (Inflation only 
fixes the sum of $\Omega_{EX}$ and $\Omega_{M}$ but not their time 
varying ratio.) Thus either way we would now have to be living in a very 
special universe, regardless in fact of what particular astrophysical form the 
currently poorly understood $\Omega_{EX}$ actually takes, with the 
evolution equation of classical cosmology thus needing some new form or other 
of fine tuning. Moreover, it is important to stress, that this 
particular classical macroscopic fine tuning problem exists independent of the 
longstanding microscopic cosmological constant problem, since it is a problem 
for the time evolution of classical cosmology rather than for the sign and 
magnitude of the vacuum energy density $\Omega_{V}$ of quantum cosmology. Thus, 
if the long sought but still as yet unidentified mechanism which supposedly 
brings $\Omega_{V}$ down from Planck density does so by actually making 
$\Omega_{V}$ vanish, an $\Omega_{EX}$ composed of something else would then 
need to be fine-tuned (by up to possibly 60 orders of magnitude); whereas if 
$\Omega_{V}$ is in fact to be the requisite $\Omega_{EX}$, a current era 
$\Omega_{V}$ of order one would then require an even greater degree of
fine tuning (120 orders of magnitude no less). To resolve this issue it thus 
becomes extremely urgent for candidate quantum theories of Einstein gravity to 
definitively ascertain the sign and magnitude of $\Omega_{V}$ once and for 
all, with the very relevance of these theories to the real world in a sense 
resting on their doing precisely that.

Beyond these issues, there is yet another concern 
\cite{Mannheim1994,MannheimandKazanas1994,Mannheim1996} 
for standard gravity, namely, 
that despite the huge amount of attention given to it for such a long time now, 
it has so far only been shown to be sufficient to account for gravitational 
phenomena, and has yet to be shown to also be necessary (and thus unique); with  
the core gravitational information garnered from study of the solar 
system serving only to show us that gravity is indeed a metric theory, and that 
on solar system distance scales the first few perturbative, weak gravity metric 
terms (the only ones so far measured) agree with those of the Schwarzschild 
metric solution to the Einstein equations, and moreover, agree with the 
Schwarzschild metric as calculated using known luminous matter alone. Then, 
as soon as this same theory is applied on much larger distance scales (from 
galaxies all the way up to cosmology) it is actually found to fail drastically 
if restricted to established visible sources. To save the situation, the theory 
is then fixed up by assuming the presence of dark matter in just the amount 
needed. (Since the amount of dark matter needed is only determined after the 
fact, dark matter cannot yet be categorized as being a falsifiable theory.) 
Since there is not yet a single independent verification of standard gravity on 
these larger distance scales which does not involve an appeal to dark matter, 
we thus see the complete circularity of the reasoning which led to dark matter 
in the first place, with such dark matter potentially being nothing more than 
an artifact which serves to parameterize any detected departure from the 
luminous 
Newton-Einstein expectation. Taken together with the implications of the new 
cosmic repulsion data described above, we identify an uncomfortably large 
number of problems for standard gravity on galactic and larger distance scales, 
and recognize \cite{Mannheim1998a,Mannheim1998b} that all of them actually 
have one single and common origin, namely the a priori use of the Einstein 
equations themselves on distance scales altogether larger than the solar system 
one where the standard theory was first established. Thus solar system, weak 
gravity Schwarzschild metric based intuition may simply not be a reliable guide 
as to the structure of gravity on altogether larger distance scales, with the 
Einstein equations simply giving an inappropriate extrapolation.

Motivated to thus look for an alternative to standard gravity with a 
potentially different large distance extrapolation, we note first that the 
covariance principle itself entails only that the gravitational action be a 
general coordinate scalar. Consequently, this action can readily be based on 
derivatives of the metric higher than the second order ones considered in the 
standard theory. For such theories curvature remains the essential ingredient, 
with only the equations used to determine the curvature being 
changed. Since the vanishing of the Ricci tensor entails the vanishing of its 
derivatives as well, higher order theories of gravity immediately both contain 
Schwarzschild (with standard gravity thus explicitly being seen to be only 
sufficient to give Schwarzschild, but not in fact necessary) and generalize it.
Moreover, the attendant departures from Schwarzschild in such higher order 
theories are precisely found 
\cite{Mannheim1994,MannheimandKazanas1994,Mannheim1996} to occur at large 
distances, i.e. at precisely those distances where the standard theory 
has to resort to dark matter, with higher order theories thus providing a very
different large distance extrapolation. While the class of contemplatable 
higher order theories is very large, it was noted \cite{Mannheim1990} that one 
of them could be singled out as special, namely a fourth order conformal 
invariant one based on the Weyl tensor, since it possesses an explicit symmetry 
(conformal invariance) which immediately sets any fundamental cosmological 
constant to zero, while simultaneously constraining any induced one to be of 
the same order of magnitude as the energy density of ordinary matter - to thus 
naturally yield an $\Omega_{V}$ of order $\Omega_{M}$ rather than one 120 
orders of magnitude larger. Conformal gravity is thus recognized as being
a theory in which the cosmological constant is actually under control, with no 
fine tuning then being required for it. In this theory the most general 
metric exterior to a static, spherically symmetric source such as a star is 
found \cite{MannheimandKazanas1994} to be given by 
$d\tau^2=B(r)c^2dt^2-B(r)^{-1}dr^2-r^2d\Omega$ where $B(r)
=1-2\beta^{*}/r+\gamma^{*} r$, to thus indeed enable us to both recover 
Schwarzschild and depart from it at large distances; with these new linear 
potential terms which are to now accompany the Newtonian potential 
actually even being found \cite{Mannheim1997} to provide for a 
successful accounting of galactic rotation curve systematics 
without the need to introduce any galactic dark matter. Further, as a 
cosmology, conformal gravity has an evolution equation different from 
Friedmann, one which possesses \cite{Mannheim1992,Mannheim1995}  
no flatness, horizon or age problems, and which is released from 
needing to contain overwhelming amounts of cosmological dark matter. So 
again a cosmological fine tuning problem is avoided, with conformal 
cosmology even being found \cite{Mannheim1992,Mannheim1997} to naturally 
have explicitly negative spatial curvature, and thus actually be far 
from flat. Finally, this same $k<0$ cosmology has even been 
shown \cite{Mannheim1998a,Mannheim1998b} to naturally produce the recently 
detected cosmic repulsion, with a $k<0$ geometry acting like a 
diverging refractive medium, to thus naturally cause galaxies to accelerate 
away from each other. 

We thus see that conformal gravity can readily address a whole host of 
issues which trouble the standard theory, viz. the cosmological constant 
problem, the flatness problem, the horizon problem, the dark matter problem, 
the universe age problem, and the cosmic repulsion problem. And regardless of 
the ultimate fate of conformal gravity itself (an appraisal of its current 
status is given in the references below), the very existence 
of all of these problems could be a warning that the extrapolation of standard 
gravity from its solar system origins all the way to cosmology might be a lot 
less reliable than is commonly believed. This work has been supported in part 
by the Department of Energy under grant No. DE-FG02-92ER40716.00.

\end{document}